\documentclass[twocolumn,showpacs,superscriptaddress,amsmath,amssymb,floatfix,apl]{revtex4-2}

\usepackage[pdftex]{graphicx}
\usepackage{dcolumn}
\usepackage{bm}
\usepackage{color}
\usepackage{xcolor}
\usepackage[shortlabels]{enumitem}

\begin{document}

\title{Spontaneous chemical reactions between hydrogen and oxygen in nanobubbles}

\author{V. B. Svetovoy}
\email[Corresponding author: ]{v.svetovoy@utwente.nl; svetovoy@yandex.ru}
\affiliation{A. N. Frumkin Institute of Physical Chemistry and Electrochemistry, Russian Academy of Sciencies, Leninsky prospect 31 bld. 4, 119071 Moscow, Russia}

\begin{abstract}
Bulk nanobubbles (NBs) generated electrochemically by short voltage pulses of alternating polarity behave differently from those produced by regular methods. Only bubbles smaller than $200\;$nm are formed in the process and their concentration is very high. Moreover, the bubbles containing both H$_2$ and O$_2$ gases disappear quickly via the combustion reaction, although the reaction in such a small volume cannot happen according to the classical combustion theory. Experimental facts about these unusual NBs are reviewed and current understanding of the observed phenomena is provided. Visualisation methods of a cloud of NBs above the electrodes are briefly discussed. Experimental signatures demonstrating the reaction between the gases in NBs are considered. A surface-assisted mechanism proposed for the combustion reactions in restricted volumes with a high surface-to-volume ratio is discussed. It it explained how the same mechanism may describe the audible explosion of microbubbles, that is observed in certain circumstances.
\end{abstract}

\maketitle


\section{Introduction}\label{sec1}

Nanobubbles have attracted significant attention in the last 20 years \cite{Seddon2012,Lohse2015,Alheshibri2016}. This attention is driven by a fundamental problem: NBs live much too long in comparison with the prediction given by the diffusion dissolution theory \cite{Epstein1950,Ljunggren1997}. On the other hand, exciting applications of NBs in environment \cite{Gurung2016,Temesgen2017}, agriculture \cite{Ebina2013}, flotation \cite{Etchepare2017,Azevedo2019}, healthcare \cite{Mondal2012,Modi2014}, and chemistry \cite{Atkinson2019} have emerged to name only a few.

In contrast with surface NBs, which exist on the solid-liquid interface, bulk NBs fill the volume of liquid. These NBs are generated by mostly mechanical injection of different gases \cite{Etchepare2017} or by ultrasonic cavitation \cite{Azevedo2019}, but hydrogen and oxygen NBs where also generated by DC electrolysis \cite{Kikuchi2007,Zhu2016}. Later these methods will be called the regular methods. The size distribution of the NBs was characterized by the dynamic light scattering and individual bubbles were observed by electron microscopy from freeze fracture replicas \cite{Ohgaki2010}, by phase microscopy and polarimetric scatterometry \cite{Bunkin2012}.

Unexpectedly, NBs containing oxygen are able to produce reactive oxygen species which can oxidize pollutants and pathogens in water \cite{Liu2013,Liu2016a,Liu2016b}, while separate oxygen molecules have no effect on organics.  Generation of OH radicals from shrinking air microbubbles (MBs) without external dynamical stimuli such as ultrasound or high pressure differential was observed by Takahashi \textit{et al.} \cite{Takahashi2007} with the electron spin-resonance spectroscopy. This finding was confirmed through reactivity with probe molecules by other researches and it was established that only OH radicals can be produced from air, O$_2$, or O$_3$ nanobubbles \cite{Liu2016a}. Observation of free radicals is a puzzling phenomenon since radical formation is a high energy event, which is not possible without external energy supply. Similar to the long lifetime of NBs the controversy on the radical production by NBs persists.

One more puzzling phenomenon that can proceed only in a high energy environment  was observed in NBs containing mixture of H$_2$ and O$_2$ gases \cite{Svetovoy2011,Svetovoy2014}. To ignite a series of combustion reactions between these gases in a fixed volume one has to produce a certain number of free radicals. Moreover, the reactions will not be self-supported if the volume is too small. This is because the heat produced by the exothermic reaction  escapes too fast via the volume boundaries and the temperature inside of the volume will be too low to support the reaction. Nanobubbles seem especially unfavorable to support the combustion reactions but in spite of this the combustion of the gases proceeds in NBs spontaneously (without ignition).

Nanobubbles containing hydrogen, oxygen, but also a mixture of the gases were generated in a so-called alternating polarity (AP) electrochemical water decomposition process \cite{Svetovoy2011} when the polarity of the electrodes is changed with a frequency of the order of $100\:$kHz or higher. In contrast with the DC electrolysis  \cite{Kikuchi2007,Zhu2016}, in this process only NBs are formed, which do not scatter the visible light but change the refractive index of liquid at a level that can be easily observed optically.

In this paper we describe briefly formation and observation of NBs in the AP process, but the main attention is directed to combustion of hydrogen and oxygen in NBs. The latest progress in the observation and understanding of the combustion reactions in NBs and closely related phenomenon observed in MBs is discussed.

\section{Signatures of the reaction between H$_2$ and O$_2$ gases in NBs}\label{sec2}

\subsection*{Generation of NBs in the AP process}

 Water electrolysis performed by short AP voltage pulses demonstrated very unusual properties in comparison with the DC electrolysis \cite{Svetovoy2011}. If one applies such pulses to the electrodes, for which the polarity of the working electrode changes with a frequency of the orders of $100\;$kHz and the opposite electrode is grounded, visible production of gas suddenly disappears  \cite{Svetovoy2011}. The transition occurs at frequencies above $20\;$kHz but the current through the electrode only increases gradually with the frequency increase. For Pt electrodes one can separate the Faraday component in the current \cite{Svetovoy2013}, which shows that the gas has to be produced with the increasing rate for higher frequencies. It was concluded \cite{Svetovoy2011} that in the AP process only NBs with a size smaller than $200\;$nm are formed. Since such bubbles practically do not scatter the visible light, they become invisible in optical microscope.

\begin{figure}[tb]
\begin{center}
\includegraphics[width=0.5\textwidth]{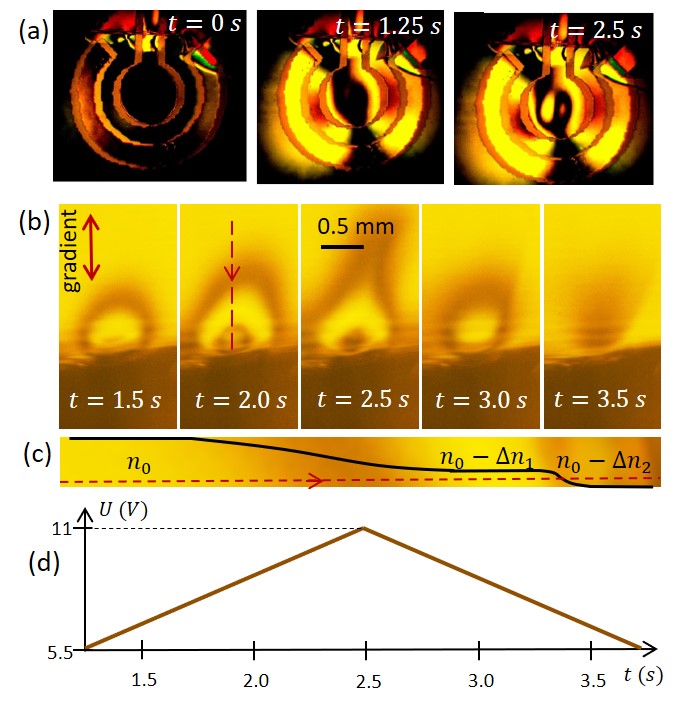}
\caption{Cloud of NBs. (a) Differential interference contrast in dynamics. The AP pulses at $f=400\;$kHz are modulated by triangle with a period of $5\;$s and with the maximum amplitude $U_{max}=8\;$V. Images for the beams of different polarization are shifted for $30\;\mu$m. (b) Schlieren contrast of the cloud in dynamics (side view) at $f=200\;$kHz and $U_{max}=11\;$V. (c) Interpretation of the contrast along the red dashed line in (b) in terms of the refraction index $n$. (d) Amplitude of the pulses as a function of time. }\label{fig:fig5}
\end{center}
\end{figure}

This conclusion was confirmed with different optical methods in later works. Although it is not possible to see the separate bubbles optically, it was rather straightforward to observe collective effects produced by NBs. To localise generation of NBs,  concentric planar electrodes with the external diameter $500\;\mu$m deposited on the oxidized Si substrate have been used. A few different methods demonstrated the presence of a dense cloud of NBs above the electrodes. Figure \ref{fig:fig5}(a) shows the differential interference contrast \cite{Postnikov2017} of two mutually coherent orthogonally polarized beams, which are slightly displaced spatially at the sample plane. The interference of the beams depends on the optical path difference in the direction of the displacement. The method is appropriate for visualisation of the gas distribution in dynamics. One can see that the cloud of NBs covers the electrode and the gradient of the refractive index becomes larger with the increase of the pulses amplitude. This method gives, however, only a qualitative picture.

Direct observation of optical distortion of the electrodes allows one to estimate the gas concentration in the liquid. Analysis performed in \cite{Postnikov2017} gave the reduction of the refractive index of about 0.18. Any impurity in the solution or Joule heating provided by the current cannot explain such a large value. The only reasonable explanation is the gas in the liquid with the concentration $n_g\approx 3.5\times 10^{26}\;$m$^{-3}$. This gas cannot stay in the dissolved state since its concentration is hundreds of times larger than the saturated value for both gases. Therefore, the gas has to be collected in NBs, which are able to accept large number of gas molecules due to a high Laplace pressure. For this $n_g$ the concentration of NBs with a radius of $40\;$nm is estimated as $n_{NB}\approx 1.5\times 10^{21}\;$m$^{-3}$. This value corresponds to an average distance between the bubbles of $90\;$nm that is only $10\;$nm larger the minimum distance between the centers of two bubbles. It means that the state of matter in the cloud can be considered as a nanofoam.

The vertical structure of the cloud was analysed with a modified schlieren method \cite{Postnikov2018} that is sensitive to the gradient of the refractive index. The AP pulses modulated by a triangle (see Fig.~\ref{fig:fig5}(d)) were applied to the electrodes. A series of corresponding schlieren images is shown in Fig.~\ref{fig:fig5}(b). The stripe (c) shows distribution of the refractive index along the red dashed line in the image (b) at $t=2.0\;$s. It corresponds to the highest concentration of NBs above the central electrode, which gradually decreases with the increase of the distance from this electrode. The total size of the dense part of the cloud is estimated as $1\;$mm. Schlieren contrast can also be generated by liquid heating. A special heater that mimics the shape of electrodes and dissipates the same power has been fabricated. The contrast from this heater has completely different character and different dynamics that is determined by internal convection rather than the amplitude of the current.

The size of NBs in the cloud was determined by the dynamic light scattering method. The signal is related to the presence of the AP pulses. At frequency $f=150\;$kHz particles with a size of $80\pm 10\;$nm have been found. At higher frequency $f=325\;$kHz the size was $60\pm 10\;$nm, however, it was not possible to make a convincing conclusion about frequency dependence. Theoretically it is expected that for higher frequencies the bubbles have to be smaller.

\subsection*{Evidence of combustion of gases in NBs}

The electrochemical decomposition of water by DC current is used to drive different microfluidic devices \cite{Kim2018}, but these devices are known to be slow due to long recombination time of the gases. This time is still within minutes even if Pt coated metal foam is used as a catalyst of the reaction between H$_2$ and O$_2$ gases \cite{Yi2015}. Slow recombination of the gases is due to the fact that under normal conditions a spontaneous reaction between hydrogen and oxygen does not occur because of a high energy barrier \cite{Lewis1987,Azatyan2010,Wang2013}. On the other hand, the reaction in MBs cannot be ignited because the heat produced by this exothermic reaction escapes too fast via the volume boundary to support the combustion \cite{Veser2001,Fernandez2002}. Experimentally the minimum bubble, where it was possible to ignite the reaction between oxygen and acetylene, was $2\;$mm in diameter \cite{Teslenko2010}.

\paragraph{Gas balance}

\begin{figure*}[tb]
\begin{center}
\includegraphics[width=1.0\textwidth]{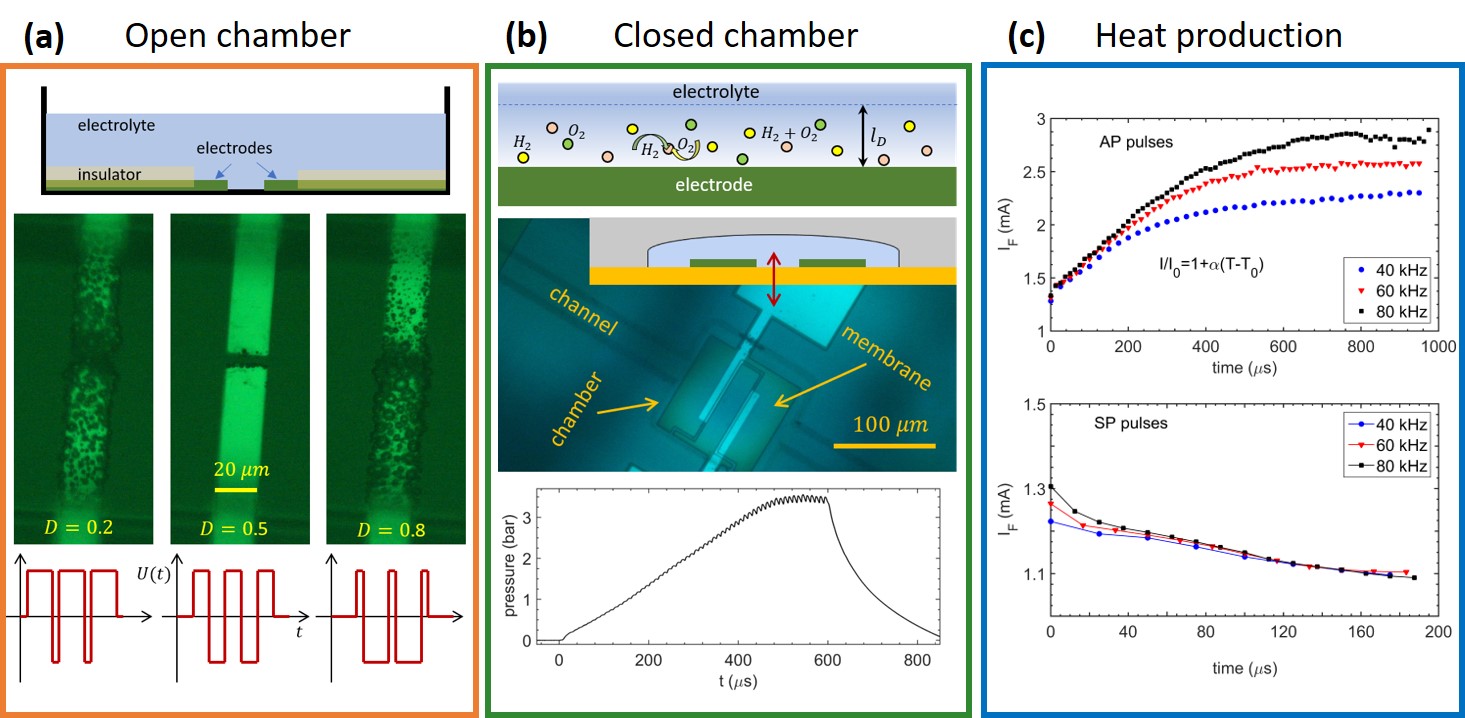}
\caption{(a) Scheme of the experiment in the open chamber (top). Images of the electrodes driven by the AP pulses with the amplitude $U=4.5\;$V and frequency $f=100\;$kHz at the moment $t=200\;\mu$s (middle). Scheme of the voltage applied to the working electrode for different duty cycles $D$ (bottom). (b) Scheme of the NBs nucleated above the electrode in the AP process (top). Image of a microfluidic chip with the closed chamber (middle). The inset shows the cross section of the chamber. Measured pressure in the chamber as a function of time (bottom). The process is driven by the AP pulses at $U=10\;$V and frequency $f=50\;$kHz during $t=600\;\mu$s. (c) The Faraday current as a function of time for the AP pulses at $U=9\;$V and for different frequencies (top). The same for SP pulses at $U=8\;$V (bottom).  }\label{fig:fig1}
\end{center}
\end{figure*}

Nevertheless, it was found that the reaction can be ignited spontaneously in very small bubbles (nanobubbles) \cite{Svetovoy2011}. As was already mentioned, generation of the gas by the AP pulses (see Fig.~\ref{fig:fig1}(a)) is proven by the Faraday current and reduction of the refractive index nearby the electrodes. If the durations of the positive and negative pulses coincide (duty cycle $D=0.5$) stoichiometric mixture of gases is produced above the same electrode. Since MBs are not observed, this gas forms NBs containing H$_2$, O$_2$, or mixture of both gases. If the average potential is shifted to the positive ($D=0.2$) or negative ($D=0.8$) side, one of the gases prevails and MBs appear in the system. Additionally, it was noted that for  $D=0.5$ the refraction index oscillates with the period equal to that of the driving pulses. It was interpreted as spontaneous reaction between H$_2$ and O$_2$ gases in NBs.

This information was confirmed and refined further in the subsequent works. To exclude the gas escape to atmosphere, the process was performed in a closed chamber just $5\;\mu$m high \cite{Svetovoy2014} (see Fig.~\ref{fig:fig1}(b)). Platinum electrodes were deposited on a thin ($150\;$nm) silicon nitride membrane that deflected proportionally to the pressure in the chamber. The pressure increased linearly with the process time and then reached a saturation state. As one can see the pressure in this state oscillates with the frequency equal to that for the driving pulses. When the pulses are switched off the pressure decreases very fast in comparison with what one would expect if no reaction between gases proceeds. It has to be stressed that at any stage of the process the MBs are not formed in the chamber.

At the top of Fig.~\ref{fig:fig1}(b) the processes happening near the surface of the electrode are shown schematically. The diffusion layer above the electrode with the thickness $l_D\sim \sqrt{tD}\sim100\;$nm is highly supersaturated with both gases, where the diffusion coefficient for the gas in liquid is $D\sim 10^{-9}\;$m$^2$/s and the timescale $t=1/f \sim 10\;\mu$s. Simple estimates show \cite{Svetovoy2014} that the relative supersaturation in this layer can exceed 1000. In this case the bubbles in the diffusion layer can be nucleated homogeneously or near homogeneously (with a low barrier) instead of growing at specific points on the electrode. Three types of the bubbles can be formed: those containing H$_2$, O$_2$, or mixture of the gases. The latter are formed and disappear each driving period that is reflected in the pressure oscillations with the frequency $f$. Moreover, the amplitude of oscillation of the refractive index above the electrodes is considerably larger than that between the electrodes. It means that most of the bubbles with the mixture of gases are created and terminated in the diffusion layer above the electrodes. The bubbles containing only H$_2$ or O$_2$ gas cannot grow larger than $l_D$ and they are pushed out to the bulk of the liquid by new bubbles nucleated in the layer. In such a way the volume of the chamber is filled with  H$_2$ and O$_2$ nanobubbles, which produce pressure resulting in the deflection of the membrane. Occasionally the bubbles with different gases merge and disappear in the reaction. When saturation state is reached in the chamber, the number of bubbles produced by the current and the number of bubbles disappearing in the reaction are equal.

It is interesting to see the gas balance in the process demonstrated by the graph in Fig.~\ref{fig:fig1}(b). The total number of gas molecules produced by the AP process during $600\;\mu$s was estimated \cite{Svetovoy2014} from the Faraday current as $N\approx 2.5\times 10^{13}$. On the other hand, the number of gas molecules in the steady state $N_{ss}$ can be estimated from the relation
\begin{equation}\label{eq:Nss}
  (P_a+\Delta P+P_L)\Delta V=N_{ss}kT
\end{equation}
that is the gas law for the gas in NBs.  Here $P_a$ is the atmospheric pressure, the overpressure in the chamber is  $\Delta P=3.6\;$bar, $P_L=36\;$bar is the Laplace pressure in a bubble with the radius $r=40\;$nm, and $kT$ is the temperature in energy units. The increase in the volume of the chamber due to the overpressure $\Delta V\approx 0.7\times10^4\;\mu$m$^3$ is taken from the collected data. This volume is equal to the total volume of NBs. From this equation one finds  $N_{ss}\approx 6.9\times 10^{12}$. It means that only 28\% of the produced gas molecules left in the steady state but the rest 72\% are consumed by the reaction. The gas concentration in the steady state is estimated as $n_{ss}=N_{ss}/(V_0+\Delta V)\approx 1.2\times 10^{26}\;$m$^{-3}$ and the concentration of NBs is $n_{NB}=(3\Delta V/4\pi r^3)/(V_0+\Delta V)\approx 0.5\times 10^{21}\;$m$^{-3}$, where $V_0=5\times10^4\;\mu$m$^3$ is the volume of the chamber. The last concentration can be compared with that for NBs produced by the regular methods $n_{NB}\sim 10^{15}\;$m$^{-3}$ \cite{Azevedo2019}.

\paragraph{Heat produced by the reaction}

 The overall reaction between hydrogen and oxygen produces significant amount of heat ($242\;$kJ/mol) that can be directly sensed. It is very important to separate the heat produced by the reaction from that generated by the current flowing through the electrolyte (Joule heat). For open systems as in Fig.~\ref{fig:fig1}(a) the effect of heating is observable but weak \cite{Svetovoy2011}. It is because the heat produced by the reaction near the microelectrodes dissipates efficiently in the substrate and thick layer of liquid. Nevertheless, the effect was measured with a resistive sensor. More detailed information on the heat flux produced by the reaction was collected with the use of a specially designed chip \cite{Jain2016}. In addition to the electrodes the chip contained a built-in resistive heat sensor insulated from the electrolyte by a thin SiN layer and a built-in heater used to relate the signal from the sensor with the heat flux going through the sensor. The integrated structure of the chip allowed accurate measurement of the heat flux. It was confirmed independently from \cite{Svetovoy2011} and \cite{Svetovoy2014} that the heat is produced by the AP process at frequency of the pulses higher than $15\;$kHz and increases with frequency up to a tested maximum of $500\;$kHz; the heat flux is reduced if the duty cycle $D$ of the pulses deviates from 0.5 in any direction. For the electrodes used in \cite{Jain2016} it was found for the heat flux $J=8\times 10^4\;$W/m$^2$ while the theoretical value expected for the stoichiometric mixture of H$_2$ and O$_2$ gases is $1\times 10^5\;$W/m$^2$. It is a good agreement for such a delicate experiment.

Although for the closed chamber it was not possible to measure the heat flux, the thermal effect of the reaction was demonstrated very clearly \cite{Svetovoy2014}. The membrane and a thin layer of liquid have a very small thermal mass in comparison with that for the bulk materials and the temperature rise due to the reaction is more pronounced. It was noted that one can use the thermal dependence of the Faraday current for the temperature sensing in the same way as for the resistive sensors. Since mobility of ions increases with temperature, the Faraday current in the electrolyte depends on the temperature as
\begin{equation}\label{eq:current}
  I_F=I_{0F}\left(1+\alpha\Delta T\right),\ \ \ \Delta T=T-T_0,
\end{equation}
where $I_{0F}$ is the current at $T=T_0$. For $1\;$M solution of Na$_2$SO$_4$ in water it was found that $\alpha=0.024\;$K$^{-1}$. Figure \ref{fig:fig1}(c) shows the Faraday current as a function of the process time for the AP pulses (upper panel) and for single polarity (SP) pulses (lower panel). The process driven by the SP pulses generates only one gas above each electrode and demonstrates decrease of the current with the time increase. It is explained by partial coverage of the electrodes with MBs and shows that the Joule heating is less significant than the effect of coverage. No frequency dependence is observed for the SP pulses. In contrast, the AP process generating the heat due to combustion reactions demonstrates the increase of the current with time and its subsequent saturation. The temperature increase depends on the frequency and can be as high as $40^{\circ}\;$C. With the increase of the driving frequency the diffusion layer becomes more homogeneous and smaller number of H$_2$ and O$_2$ bubbles leave this layer. On the other hand more bubbles with mixture of the gases are formed producing more heat.

\bigskip
Let us summarize the signatures of the reaction between hydrogen and oxygen in NBs. (i) Periodic reduction of the refractive index of liquid and pressure (in closed chamber) with the period equal to the driving period. (ii) Steady state for the pressure in the closed chamber demonstrating the balance between produced and reacted gas. (iii) Fast relaxation of the pressure in the chamber after switching off the driving pulses. (iv) Heat produced by the reaction and observed by different methods.

\section{Mechanism of the reaction}\label{sec3}

\paragraph{Domination of the interface}
Due to a very large surface-to-volume ratio for NBs, the heat escapes too fast from such bubbles. The time for heat dissipation from a bubble with the radius $r$ is $t\sim (r^2/\chi_l)(C_{pg}/C_{pl})^2\sim 10^{-13}\:$s, where $\chi_l\sim 10^{-7}\;$m$^2$/s is the heat diffusion coefficient in water and $C_{pg}/C_{pl}\sim 10^{-3}$ is the ratio of the heat capacities in gas and liquid states. This time is actually shorter than the reaction time $t_r\sim 10\;$ns (see below). Therefore, the temperature in the bubble cannot be high. It is also supported by the observations \cite{Svetovoy2011} where no hot spots have been found with a high sensitivity camera. In this case, the interface between the gas and the liquid has to play a special role in the reaction mechanism.

Bubbles in water \cite{Takahashi2005,Creux2007} and similar oil drops in water \cite{Marinova1996,Beattie2004} carry a negative charge. The $\zeta$-potential of both increases with pH from zero at $\textrm{pH}=2-4$ up to $\zeta=-120\;$mV at $\textrm{pH}=10$. The surface density of charges measured for oil drops at neutral pH is $n_s=(3-4)\times10^{13}\;$cm$^{-2}$ and a similar value is expected for the bubbles \cite{Beattie2009}. The $\zeta$-potential of the bubbles and drops is associated with hydroxyl ions adsorbed on the interface. Not all authors support this point of view \cite{Vacha2012}, but the presence of the negative charges is not disputed.

\begin{figure}[tb]
\begin{center}
\includegraphics[width=0.5\textwidth]{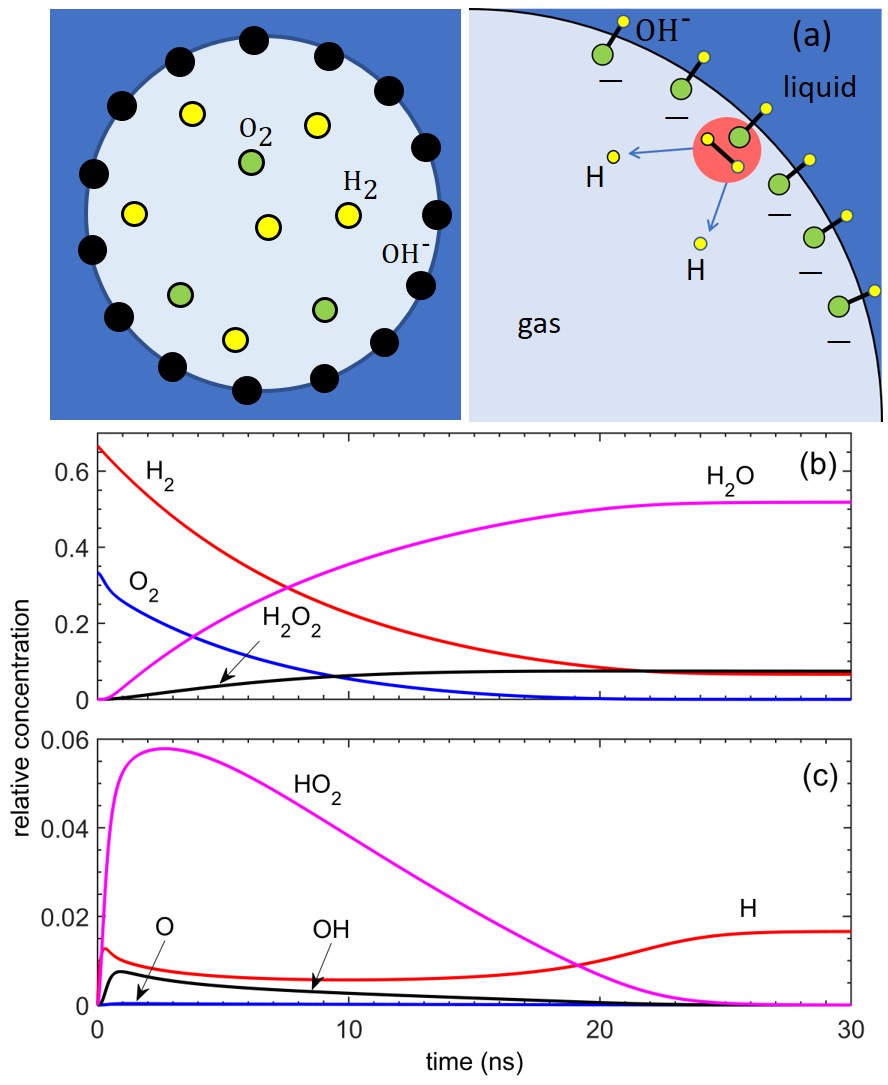}
\caption{(a) Scheme of the spontaneous ignition of the combustion in a NB with charged interface. It is assumed that molecules approaching the reaction center at the interface are able to produce radicals. (b) Solution of the equations of chemical kinetics for long-lived species in the process of surface-assisted (cold) combustion. (c) The same for short-lived radicals. All concentrations are shown with respect to the initial gas concentration.}\label{fig:fig2}
\end{center}
\end{figure}

A mechanism of the surface-assisted combustion was proposed \cite{Prokaznikov2017}, although it is able to explain the puzzle only partially. Generation of OH radicals by shrinking microbubbles observed in \cite{Takahashi2007,Li2009} was used as a hint. If these radicals can be generated in NBs with a sufficiently large surface-to-volume ratio, then one could expect generation of other radicals on the gas-liquid interface. The surface charges can assist in the formation of radicals, but the exact mechanism of this assistance is still unknown. It was postulated \cite{Prokaznikov2017} that the collision of an H$_2$ or O$_2$ molecule with the charged interface with some probability can generate H or O radicals (see Fig.~\ref{fig:fig2}(a)). The surface reactions seem to be the only way to explain spontaneous combustion in a small volume at room temperature.

The reaction constant at the surface explicitly depends on the surface-to-volume ratio $S/V$ \cite{Wang2013}
\begin{equation}\label{eq:r_const}
  K_i=(\varepsilon_i/4)\bar{v}_i(S/V)
\end{equation}
where $\bar{v}_i$ is the average thermal velocity for the $i$-species. The parameter $\varepsilon_i$ can be considered as the probability of the surface reaction. This probability can be presented as $\varepsilon_i=\sigma_in_s$, where $\sigma_i$ is the cross-section for $i$-th radical formation on the surface and $n_s$ is the concentration of the active centers on the bubble wall. Taking $n_s\sim 10^{13}\;$cm$^{-2}$ and a typical cross-section $\sigma\sim 1\;${\AA}$^2$ one finds the expected probability of the surface reactions as $\varepsilon\sim 10^{-3}$.

\paragraph{Path of the reaction}

In the classical combustion process the chemical branching reactions $\textrm{H}+\textrm{O}_2 \rightarrow \textrm{O}+\textrm{OH}$ and $\textrm{O}+\textrm{H}_2 \rightarrow \textrm{H}+\textrm{OH}$ play a key role \cite{Lewis1987}. These reactions are characterized by high energy barriers and they will be strongly suppressed if the temperature in the bubble is not high enough. Therefore, the classical combustion in NBs is not possible since the heat escapes too fast from small bubbles. On the other hand, according to (\ref{eq:r_const}) the role of the surface reactions increases with the increase of the ratio $S/V$. The radicals formed on the surface trigger a chain of the low-energy reactions transforming H$_2$ and O$_2$ gases into H$_2$O. The equations of chemical kinetics for four radicals H, O, OH, HO$_2$ and four molecules H$_2$, O$_2$, H$_2$O, H$_2$O$_2$ have been solved \cite{Prokaznikov2017}. In general 19 low-energy reactions in the bulk (with the reaction constant $>1\;\mu$s$^{-1}$) were included in the analysis and their reaction constants were collected from different sources.

The most important conclusion is that there exists a chain of low-energy bulk reactions that transforms mixtures of H$_2$ and O$_2$ gases with the assistance of the surface processes into water. For the probability of hydrogen radical generation at the surface $\varepsilon_H=0.003$ the time evolution of different species is shown in Fig.~\ref{fig:fig2}(b),(c) starting from the stoichiometric mixture of H$_2$ and O$_2$ gases. The characteristic process time $t_r\sim 10\;$ns is defined by the reaction constants. In contrast with the standard combustion there is in excess of hydrogen peroxide in the final state. Some amount of H$_2$ and H radicals in the final state is because the slow reactions (for example, decomposition of H$_2$O$_2$) were not included in the analysis. Generation of H radicals on the surface is principal for the surface-assisted combustion, but generation of O radicals does not play a key role although it can influence on the amount of hydrogen peroxide in the final state. Among the bulk reactions the principal role belong to the trimolecular reaction $\textrm{H}+\textrm{O}_2+\textrm{M}\rightarrow \textrm{H}\textrm{O}_2+\textrm{M}$ where M is any third species taking part in the process. The surface-assisted combustion becomes impossible if this reaction is excluded. A significant amount of hydrogen peroxide in the final state is related to the reaction $\textrm{H}\textrm{O}_2+\textrm{H}\textrm{O}_2\rightarrow \textrm{H}_2\textrm{O}_2+\textrm{O}_2$. The data in Fig.~\ref{fig:fig2} are presented for a bubble radius of $50\;$nm. If the radius of the NB decreases, the amount of  H$_2$O$_2$ in the final state increases.

\paragraph{Molecular dynamics simulation}

Independently the surface-assisted combustion has been analysed \cite{Jain2018} using the reactive molecular dynamics simulation. To simplify the calculations a cubic volume filled with the gases was considered instead of a sphere, but the extra pressure in this volume was introduced to account for the Laplace pressure. The method used avoids the restrictions of continuous media approach; it does not need such parameters as heat diffusion rates, bi and tri-molecular reaction rates, and pressure in the bubble which are not always known. Generation of radicals on the walls was modeled inserting a certain number of radicals near the walls or inserting the radicals periodically. The main conclusions of Ref.~\cite{Prokaznikov2017} were confirmed qualitatively. The quantitative agreement cannot be expected since the chemical kinetics describes the system on a relatively long timescale up to $1\;\mu$s while the molecular dynamics can describe the system on a timescale shorter than $0.1\;$ns.

It was confirmed by the direct simulation that the temperature in the bubble cannot reach the value of $700^{\circ}\;$C needed for standard combustion but it can be as high as $100^{\circ}\;$C. It does not contradict to the statement \cite{Prokaznikov2017} that temperature in NBs is close to room temperature. The temperature in the bubble is equalized during $t\sim r^2/\chi_g\sim 1\;$ns where $\chi_g\sim 10^{-6}\;$m$^2$/s is the heat diffusion coefficient in gas. This time cannot be reached by the molecular dynamics. It was confirmed also that the final products contain a significant amount of hydrogen peroxide. It was stressed that for the surface-assisted combustion the main intermediate product is HO$_2$ while for the normal combustion the most important intermediate product is OH radicals. Important role of HO$_2$ can also be seen in Fig.~\ref{fig:fig2}(c).

\bigskip
Summarizing the current understanding of the mechanism of the reactions in NBs we can separate the following important points. (i) Combustion in NBs is the low-temperature process closely related to a high surface-to-volume ratio for NBs. (ii) If H radicals can be generated at the gas-liquid interface, then there is a chain of elementary reactions that transforms stoichiometric mixture of H$_2$ and O$_2$ gases into water and hydrogen peroxide. (iii) Generation of H radicals is assumed to be related to the charges at the gas-liquid interface but the precise mechanism is still unknown.

\section{Combustion in microbubbles}\label{sec4}

\begin{figure}[t]
\begin{center}
\includegraphics[width=0.49\textwidth]{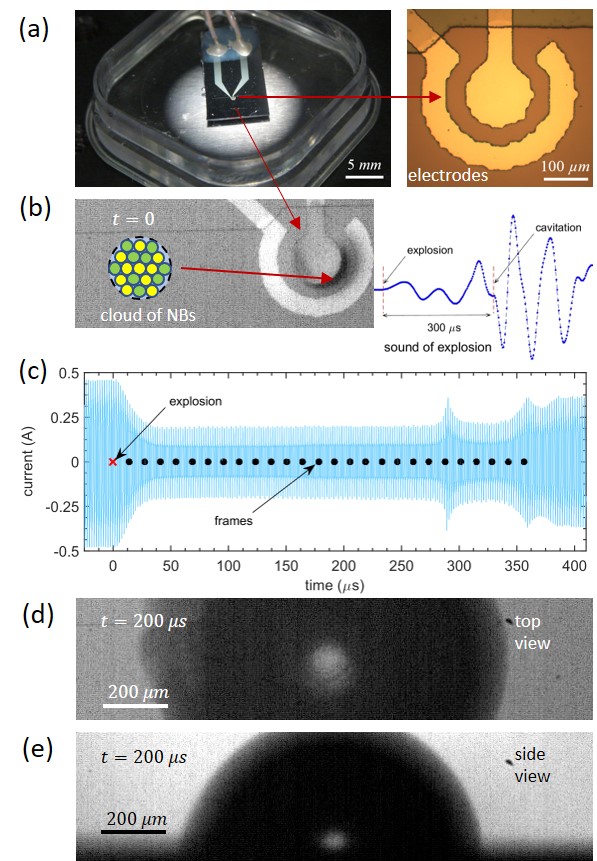}
\caption{Explosion of a microbubble. (a) The sample fixed in small Petri dish filled with Na$_2$SO$_4$ solution and zoomed view of Ti concentric electrodes. (b) Electrodes just before the explosion. The cloud of NBs is so dense that it scatters slightly the visible light. Right side shows the sound amplitude that demonstrates two sources of the sound. (c) Current between the electrodes drops synchronously with the growing MB. The AP process is at $U=14\;$V and $f=500\;$kHz. Black circles show positions of the frames in the fast video. (d) The MB in its maximum size observed from the top. (e) The same as (d) but viewed from the side.}\label{fig:fig3}
\end{center}
\end{figure}

The main conclusion from the previous section is that the surface-assisted combustion can be realised only for a high surface-to-volume ratio. All the more surprising that the combustion was observed in MBs produced at special conditions by the AP process \cite{Postnikov2016}. To investigate this phenomenon in detail the concentric electrodes shown in Fig.~\ref{fig:fig3}(a) were used \cite{Svetovoy2020}. For such electrodes the NBs are well concentrated between the electrodes.

\paragraph{Experimental facts}

With the increase of the amplitude of the pulses more and more NBs are generated in the AP process. When the concentration of NBs reaches a critical value, a very interesting phenomenon occurs. The cloud of NBs visible in the left side of Fig.~\ref{fig:fig3}(b) suddenly is transformed into a MB that explodes with a clearly audible 'click' sound. The clicks are repeated with a frequency of $12\;$Hz. The sound consists of two sources separated by a time interval of $300\;\mu$s as one can see in the right side of (b). Synchronously with the sound the current in the system decreases in just $20\;\mu$s (c) and stays low during $300\;\mu$s. Absence of a sharp increase in the current together with insufficient electrical field demonstrate that the phenomenon cannot be related to the electrical breakdown. Observation of the process with a fast camera demonstrates that synchronously with the current decrease a MB with an initial size of $150\;\mu$m appears above the electrodes, grows to a maximum size of $1200\;\mu$m in $200\;\mu$s, than shrinks fast, and disappears. The MB in its maximum is shown in Fig.~\ref{fig:fig3}(d) and (e) from the top and from the side respectively. Reduction of the current is explained by the growing bubble, which overlaps the electrodes in $20\;\mu$s.

\begin{figure*}[tb]
\begin{center}
\includegraphics[width=1\textwidth]{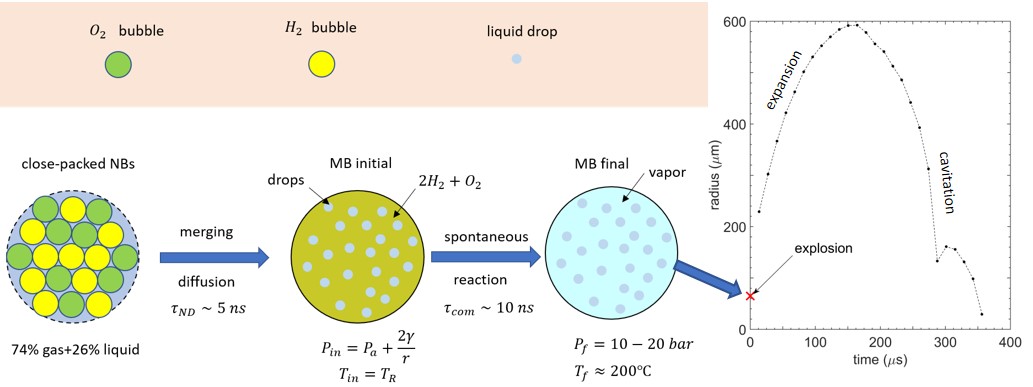}
\caption{(Left) Scheme explaining formation of the MB and chemical reaction happening in the bubble. (Right) Evolution of the final MB observed with the fast camera with an interval between the frames of $13.7\;\mu$s.  }\label{fig:fig4}
\end{center}
\end{figure*}

For understanding of the phenomenon it is important that there are two sources of the sound accompanying the process. The first sound appears together with the decrease of the current and when the current starts to restore the second more powerful sound emerges. The second sound appears in the moment when the bubble reaches its minimum value and can be explained by the cavitation effect. From the fast video one can find the time dependence of the bubble radius as shown in Fig.~\ref{fig:fig4} (right). In this dependence one can resolve the second bump that is a characteristic feature of cavitation. Moreover, after many clicks the silicon sample with the deposited Ti electrodes is destroyed locally very similar to that produced by cavitating bubbles (Figure S1 in \cite{Svetovoy2020}).

The first sound one can relate to the explosion in the initial bubble. It starts synchronously with the current decrease and obviously is related to the expanding MB. The expansion rate estimated from the fast video is about $8\;$m/s. This rate is in agreement with a typical velocity of the burning front for normal combustion. However, the observed process cannot be induced by normal combustion because the heat escapes from the bubble with a radius of $75\;\mu$m too fast ($t\sim 10\;$ns) to support normal combustion. A special experiment was performed to check the possibility of spontaneous combustion of H$_2$ and O$_2$ mixture in MBs. The stoichiometric mixture of gases was pumped into the gas channel of a microfludic bubble generator. The MBs with a diameter of $10\;\mu$m moving in water one by one was observed in the output channel. No spontaneous combustion in these bubbles was observed \cite{Postnikov2016}. It seems that the surface-assisted combustion also cannot explain the phenomenon of expanding MB because the $S/V$ ratio in Eq.~(\ref{eq:r_const}) is three orders of magnitude smaller than that for NBs.

\paragraph{Role of nanodrops}

Nevertheless, the process has been interpreted  \cite{Svetovoy2020} as the surface-assisted combustion in MBs. The main steps of this process are shown schematically in Fig.~\ref{fig:fig4}. As was already mentioned, when the amplitude of the pulses increases, the NBs approach very close to each other and the density of the cloud approaches the critical value. This value corresponds to the close packing of spheres and is equal $n_{NB}\approx 2\times 10^{21}\;$m$^{-3}$ for $r=40\;$nm.

The cloud consists of H$_2$ and O$_2$ NBs. When two similar bubbles are close to each other, there is no significant diffusion of gases since the concentration gradient is low. However, when two bubbles with different gases meet, fast exchange by the gases occurs. The minimum distance between the bubbles is defined by the disjoining pressure in the separating membrane with the thickness $d\sim 1\;$nm. Then the exchange time by the gases is estimated as $t\sim d^2/D_l\sim 1\;$ns, where $D_l\sim 10^{-9}\;$m$^2$/s is the diffusion coefficient of gas in liquid. It means that the cloud of NBs with the critical density will be transformed to a MB containing mixture of H$_2$ and O$_2$ gases. However, the volume filled with the close-packed NBs has the volume fraction of gas $f_0=0.74$ and the rest is liquid. After merging of NBs this liquid will be collected in nanodrops with the radius $r'=r[(1-f_0)/f_0]^{1/3}$.

The MB formed by merging of many NBs (initial MB) is filled with the stoichiometric mixture of gases, but 24\% of its volume is occupied by the nanodrops. The surface-to-volume ratio in this MB is $S/V=3/r'\sim 1/r$. This ratio is nearly as high as that for the NBs and for this reason it is able to support the surface-assisted reactions. These reactions can proceed in the same way as it was described in Sec.~\ref{sec3}. On a timescale of $10\;$ns hydrogen and oxygen in the MB will turn into water vapor (final MB). In spite of a large size of the MB the thermal equilibrium in the final MB is established on the same timescale as for NBs $t\sim 1\;$ns because of the presence of nanodrops separated by the average distance $l=r(4\pi/9f_0)^{1/3}\sim r$. Therefore, the final MB has well-defined pressure and temperature, which determine the following evolution of the bubble directly observed in the experiment. This evolution is much more slow (the time scale $t> 1\;\mu$s) and, in principle, can be described by the Rayleigh–Plesset equation \cite{Brennen1995} with the accurate inclusion of the heat exchange effects between the MB and surrounding liquid. Therefore, the final MB is in the state, in which the bubble has not yet expanded, but the chemical reaction is already over and the steam–liquid equilibrium is established.

\paragraph{State in the exploding MB}

Owing to the thermal equilibrium it is possible to find the pressure $P_f$ and temperature $T_f$ in the final MB using only the energy balance equation \cite{Svetovoy2020}. The internal energy of the initial MB plus the energy produced by the reaction are spent on the energies of the gas phase, liquid phase, and partial vaporization of nanodrops in the final MB. The effect of surface tension in the balance is insignificant. Since the gas and liquid phases in the final MB are in equilibrium, the pressure in the bubble is defined by the temperature $P_f=P_{eq}(T_f)$, where the function $P_{eq}(T)$ and other thermodynamic values were taken from the vapor-liquid equilibrium tables \cite{Beaton1988}. The resulting value depends on the size of merging NBs. For $r=30\;$nm it was found that $T_f=220.8^{\circ}\;$C and $P_f=23.5\;$bar but for $r=40\;$nm these parameters are $T_f=185.5^{\circ}\;$C and $P_f=11.4\;$bar.

The pressure in the final MB is much higher than that in the surrounding liquid $P_f\gg P_a$ and this pressure jump is the source of the first sound shown in Fig.~\ref{fig:fig3}(b). The temperature in the final MB is higher than that found for NBs \cite{Jain2018}, but it is still not enough for the spontaneous ignition of the normal combustion.

\bigskip
Summarizing combustion in MBs one can stress the following. (i) The process is observed only for MBs formed from a cloud of NBs produced in the AP process; no reactions happen in the MBs filled with H$_2$ and O$_2$ gases taken from external sources. (ii) Explosion of the initial MB is accompanied by the sound of explosion and after a delay of $300\;\mu$s by the sound of cavitation. (iii) Nanodrops that left in the MB after merging of many NBs provide the surface-to-volume ratio as high as in a separate NB and ensure conditions for the surface-assisted combustion. (iv) The final MB immediately after the chemical reaction is characterized by the pressure $P_f=10-20\;$bar and the temperature $T_f\approx 200^{\circ}\:$C; high pressure in the final MB is the source of the first sound.

\section{Open problems}\label{sec5}

We still do not know if the NBs generated in the AP process are the same as those produced by the regular methods. Short lifetime of the bubbles generated in the AP process is explained by the interaction between  H$_2$ and O$_2$ NBs.
However, if we menage to separate hydrogen and oxygen NBs, for example by diluting the solution, would they live much longer? A recent paper \cite{Satpute2021} relates stability of the regular NBs to their origin. It is assumed that long-lived NBs appear only from shrinking MBs and their stability is related to the increasing concentration of the surface charges. The answer to the above question will be a critical test for this hypothesis because only NBs are produced in the AP process.

It is not clear why only NBs are generated by the AP pulses. We know that very high supersaturation supports homogeneous nucleation of NBs, but for some reason these bubbles do not grow or merge easily. It is expected that such processes are controlled by the surface charges and opposite charges distributed in the electrolyte. However, these effects have never been analysed in depth.

Although the surface-assisted combustion is able to give a basic explanation of the reactions in NBs and in MBs, generation of radicals on the gas-liquid interface is the most important open question. Direct evidence for the presence of the radicals is still lacking. It could be obtained using the electron spin resonance spectroscopy or using different radical scavengers in the solution. It is also possible that generation of OH radicals in the regular NBs and H (O) radicals in NBs produced by the AP process are related effects. In this respect it is interesting to consider a reaction path for 'cold' combustion via OH radicals generated on the surface instead of H or O radicals as in \cite{Prokaznikov2017}.

Dynamics of exploding MB was determined experimentally. On the other hand, the evolution of exploding NB also has a special interest. Recently it was demonstrated \cite{Uvarov2021} that a flux of H$_2$ and O$_2$ NBs directed to a Pt plate is able to produce Pt nanoparticles that is a high energy process. Understanding of the dynamics of exploding NBs can shed light on the observed phenomena.

\section{Conclusions}\label{sec6}

In this paper we presented the current status and understanding of the phenomena observed during generation of NBs by short alternating polarity voltage pulses. In contrast with the regular methods no bubbles larger than $200\;$nm are produced by the AP method. The concentration of NBs is so high that one can easily observe collective optical effects produced by these NBs. Since high supersaturation with both H$_2$ and O$_2$ gases is generated above the same electrode, three types of NBs can be produced: those containing  H$_2$, O$_2$, and mixture of the gases.

The most important difference from the regular NBs is that inside of one NB hydrogen and oxygen are able to react with each other in spite of the expectations from the classical combustion theory. It is clear demonstrated by a series of experimental facts such as pressure oscillation in a closed volume with the frequency of the driving pulses, fast relaxation of the pressure after switching off the pulses, heat produced by the reactions, and others.

A possible mechanism of the combustion reactions in a small volume relies on a very high surface-to-volume ratio characterizing NBs. It assumes generation of free radicals on the gas-solid interface presumably due to the charges known to be adsorbed on the interface. With this assumption the chain of 'cold' reactions exists, which is able to transform mixture of hydrogen and oxygen into water without necessity of  high temperature. However, the exact mechanism of the radical formation at the interface is still unknown.

At special conditions combustion of the gases is observed in MBs that is accompanied by audible clicks, by significant  expansion of the bubble followed by the cavitation. It seems that the combustion in MB cannot be described by the surface-assisted mechanism because MBs are characterized much smaller surface-to-volume ratio. However, as described in the paper a MB formed by merging of many NBs contains nanodrops so that the total surface-to-volume ratio is of the same order as for NBs. Theoretical estimates of the pressure and temperature in the MB immediately after the reaction show that the sound and expansion are produced by a pressure jump of $10-20\;$bar but the gas temperature stays relatively low of about $200^{\circ}\:$C.

Finally we discuss different problems that need deeper understanding. The most important problem is the mechanism of radical formation at the interface.


\end{document}